\documentclass[preprint,prb]{revtex4}
\usepackage{graphicx}
\usepackage{dcolumn}
\usepackage{bm}
\newcommand{\be}{\begin{equation}}
\newcommand{\ee}{\end{equation}}
\newcommand{\bea}{\begin{eqnarray}}
\newcommand{\eea}{\end{eqnarray}}

\begin{document}

\draft
\title
{\bf Bistable states of quantum dot array junctions for high-density memory}

\author{ David M.-T. Kuo$^{1}$ and Yia-Chung Chang$^{2,3}$ }
\address{Department of Electrical Engineering, National Central
University,\\
Chung-Li, Taiwan 320, Republic of China}

\address
{$^2$Research Center for Applied Sciences, Academia
Sinica, Taipei, Taiwan 115, R.O.C.}
\address
{$^3$Department of Physics University of Illinois at Urbana-Champaign, Urbana, Illinois 61801}

\date{\today}

\begin{abstract}
We demonstrate that two-dimensional (2D) arrays of coupled quantum dots (QDs) with six-fold
degenerate p orbitals can display bistable states, suitable for application in high-density
memory device with low power consumption. Due to the inter-dot coupling of $p_x$ and $p_y$ orbitals in these QD arrays,
two dimensional conduction bands can be formed in the x-y plane, while the  $p_z$ orbitals remain localized in the x-y plane such that
the inter-dot coupling between them can be neglected. We model such systems by
taking into account the on-site repulsive interactions between
electrons in $p_z$ orbitals and the coupling of the localized
$p_z$ orbitals with the 2D conduction bands formed by $p_x$ and $p_y$ orbitals.
The Green's function method within an extended Anderson model is used to calculate the
tunneling current through the QDs. We find
that bistable tunneling current can exist for such systems due to the
interplay of the on-site Coulomb interactions (U) between the
$p_z$ orbitals  and the delocalized nature of conduction band
states derived from the hybridization of $p_x$/$p_y$ orbitals.
This bistable current is not sensitive to the detailed band
structure of the two dimensional band, but depends critically on
the strength of $U$ and the ratio of the left and right tunneling
rates. The behavior of the electrical bistability can be sustained when the 2D QD array reduces to a one-dimensional QD array,
indicating the feasibility for high-density packing of these bistable nanoscale structures.
\end{abstract}

\maketitle

Intrinsic hysteresis in DC current-voltage characteristics is one
of the most intriguing problems for resonant tunneling diodes
(RTDs)$^{1,2}$. Such a bistability has an important application in
memory devices$^{3,4}$. Whether this phenomenon exists in
nanoscale devices such as single-electron transistors (SETs) and
single molecular transistors (SMTs) has been theoretically
investigated in refs [5-7]. Alexandrov and coworkers$^{5}$ pointed
out that the tunneling current through a highly degenerate states
of a single QD (molecular) can lead to a switching effect only in the
case of attractive electron Coulomb interactions, which is
mediated by electron-phonon interaction.
On the basis of Hartree approximation and
polaron effect Galperin et al proposed that the hysteresis of I-V
characteristics can be observed in a single molecular junction
with effective attractive electron Coulomb interaction.$^{6}$
Recently,  Magna and Deretzis showed hysteresis feature of
tunneling current in a polaron model beyond the Hartree
approximation.$^{7}$

Although previous theoretical studies predicted the existence of hysteresis
in a QD (or molecular) junction,$^{5-7}$ such a phenomenon still
lacks conclusive experimental support. Moon et al. have
experimentally examined the tunneling current through a carbon
nanotube QD, which exhibits a periodic oscillatory behavior with
respect to the applied gate voltage arising from the eightfold
degenerate state.$^{8}$ In addition, Liljeroth et al. have
reported a periodic oscillatory differential conductance as a
result of tunneling current through a single spherical PbSe QD
with a sixfold degenerate state$^{9}$. These two experiments did
not exhibit the bistable tunneling current. Their results indicate
that electron-phonon interactions in nanotube QDs or PbS QDs are not
sufficient to yield the strongly attractive Coulomb interactions
needed for observing the bistability. Thus
it remains questionable whether a single QD junction can display the
bistable memory effect.

Recently, it was demonstrated that semiconductor quantum dot arrays (QDAs) can be
chemically fabricated to form a superlattice.$^{10-13}$ Via
nanoscale manipulation, experimentalists can now control the
lattice constant and QD size to tune charges of QDA in the Coulomb
blockade regime or semiconducting regime.$^{10,14}$ Consequently,
QDA is not only a good physical system for investigating strongly
correlated problem but also a promising integrated electronic
device.$^{10,15}$ Although many theoretical efforts have been
devoted to the charge transport through a single QD,$^{16,17}$ not
many studies are on the tunneling current through a QDA
junction.$^{18}$ In this letter we illustrate that a new mechanism
exists in  a QDA junction involving degenerate p-like orbitals
which can lead to bistable tunneling current, making it a good candidate for high density storage device.

Figure. 1 illustrates the system of a QD array embedded in an
insulator connected with metallic electrodes. The system can be
described by the Anderson Hamiltonian, $H=H_0+H_T+H_d$. The  $
H_0=\sum_{k,\sigma,\beta} \epsilon_k
a^{\dagger}_{k,\sigma,\beta}a_{k,\sigma,\beta}$ describes the
electronic states in the metallic leads.Here
$a^{\dagger}_{k,\sigma,\beta}$ ($a_{k,\sigma,\beta}$) creates
(destroys) an electron of momentum $k$ and spin $\sigma$ with
energy $\epsilon_k$ in the $\beta$ metallic electrode. The  $H_T$
term describes the coupling between the electrodes and the $p_z$
orbitals of the QD array.
\begin{equation}
H_T=\sum_{k,\sigma,\beta,\ell} V_{k,\beta,\ell}a^{\dagger}_{k,\sigma,\beta}d_{\ell,\sigma}\\
+\sum_{k,\sigma,\beta,\ell}
V^{*}_{k,\beta,\ell}d^{\dagger}_{\ell,\sigma}a_{k,\sigma,\beta},
\end{equation}
where $V_{k,\beta,\ell}$ describes the coupling between the band
states in the electrodes and the localized $p_z$ states. Here we
assume that the coupling between the electrodes and the $p_x/p_y$
orbitals of the QD array is negligible since the $p_x/p_y$
orbitals are much more localized along the $z$ axis than the
$p_z$ orbitals. At last, the $H_d$ term describes electronic
states and their interactions in the QD array.
\begin{eqnarray}
H_d &=&\sum_{\ell,\sigma} E_{p_z} d^{\dagger}_{\ell,\sigma}
d_{\ell,\sigma} + \sum_{p,\lambda}
(\epsilon_{p,\lambda}+U(N_c-N_\lambda)) c^{\dagger}_{p,\lambda}
c_{p,\lambda}\\
\nonumber & + & \sum_{\ell,p,\sigma}(v_{p,\ell}
c^{\dagger}_{p,\sigma} d_{\ell,\sigma}+h.c) +\sum_{\ell,\sigma} U_{\ell}
d^{\dagger}_{\ell,\sigma}
d_{\ell,\sigma} d^{\dagger}_{\ell,-\sigma} d_{\ell,-\sigma} \\ \nonumber &+&
 \frac{U_{dc}}{N} \sum_{\ell, p,p',\sigma} c^{\dagger}_{p,\lambda}
c_{p',\lambda} e^{i({\bf p-p}')\cdot {\bf R}_\ell} d^{\dagger}_{\ell,\sigma,\lambda} d_{\ell,\sigma}.
\end{eqnarray}


$d^{\dagger}_{\ell,\sigma}$ ($d_{\ell,\sigma}$) creates (destroys)
an electron in  the $p_z$ orbital (with energy $E_{p_z}=E_a$) of
the QD at site $\ell$. The second term in Eq. (2) describes the
conduction bands of QD array arising from the $p_x$ and $p_y$
orbitals. $\lambda$ labels the conduction bands (including spin).
$U$ denotes the on-site Coulomb interaction between two electrons
in the $p_x$ and $p_y$ orbitals. Note that if we ignore the
quadrupole and higher-order terms in the expansion of $1/r_{12}$,
then the Coulomb repulsion integrals between two electrons in any
of the three degenerate p-like orbitals are the same. $N_\lambda$
is the occupation number per unit cell for the $\lambda$-th
conduction band, and $N_c=\sum_\lambda N_\lambda$ is the total
occupation number per unit cell for the conduction bands. A
mean-field theory (which is justified for extended states) has
been applied to the 2D conduction bands to obtain the second term
in the above equation. The third term in Eq. (1) describes the
hopping  coupling between the $p_z$ orbital and the  $p_x/p_y$
orbitals within the tight-binding model.  The last two terms in
Eq. (2) involve $U_{\ell}=U$, and $U_{dc}$, which denote the
on-site repulsive Coulomb energy in the $p_z$ orbital, and
electron Coulomb interactions between the $p_z$ and $p_x/p_y$
orbitals. $N$ denotes the number of QDs in the matrix. Here, we
focus on the  $p_z$ orbital rather than the ground state orbital,
even though its wave function is more localized than that of
$p_z$, since in the range of applied bias considered, the QD
ground state energy level is deeply below the Fermi levels of both
electrodes and the electron tunneling through the QD ground state
is blockaded. Consequently, carriers in the QD ground states only
lead a constant-shift to all the p orbitals. It is worth noting
that Eq. (2) is similar to the so-called the extended
Falicov-Kimball model, which has been used extensively to study
the semiconductor-metal transition in a solid consisting of
localized orbitals and delocalized orbitals.$^{19-21}$


Using Keldysh Green's function technique$^{22,23}$, the tunneling
current through the $\ell$th QD can be expressed by
\begin{equation}
J_{\ell,\sigma}=\frac{-e}{\hbar}\int
\frac{d\epsilon}{\pi}[f_{L}(\epsilon)-f_R(\epsilon)]
\frac{\Gamma_{\ell,L} \Gamma_{\ell,R}}
{\Gamma_{\ell,L}+\Gamma_{\ell,R}}ImG^r_{\ell,\ell}(\epsilon),
\end{equation}
where $f_{L}= f(\epsilon-\mu_{L})$ and $f_{R}=f(\epsilon-\mu_{R})$
are the Fermi distribution functions for the left and right
electrodes, respectively. The chemical potential difference
between these two electrodes is related to the applied bias by
$\mu_{L}-\mu_{R}=eV_a$. $\Gamma_{\ell,L}(\epsilon)$ and
$\Gamma_{\ell,R}(\epsilon)$ [$\Gamma_{\ell,\beta}=2 \pi \sum_{{\bf
k}} |V_{\ell,\beta,{\bf k}}|^2 \delta(\epsilon-\epsilon_{{\bf
k}})]$ denote the tunneling rates from the $p_z$ orbitals to the
electrodes. Notations $e$ and $\hbar$ denote the electron charge
and Plank's constant. In the wide-band limit, these tunneling
rates are approximately energy-independent. Therefore, the calculation of
tunneling current is entirely determined by the spectral function
$A=ImG^r_{\ell,\ell}(\epsilon)$, which is the imaginary part of
the retarded Green's function $G^r_{\ell,\ell}(\epsilon)$.

Using the equation of motion for $G^r_{\ell,\ell}$, we obtain

\begin{eqnarray}
& &(\epsilon-E_0+i\Gamma) G^{r}_{i,j}(\epsilon)\\ \nonumber &=&
\delta_{i,j}+U<n_{i,-\sigma}d_{i,\sigma}d^{\dagger}_{j,\sigma}>\\
\nonumber &+&\sum_p v_{i,p}G^{r}_{p,j}+ \sum_{p^{''},p^{'},\sigma}
g_{p^{''},p^{'}}
<c^{\dagger}_{p^{''},\sigma'}c_{p^{'},\sigma'}d_{i,\sigma}d^{\dagger}_{j,\sigma}>,
\end{eqnarray}

\begin{eqnarray}
& &(\epsilon-E_0+i\Gamma) G^{r}_{i,p}(\epsilon)\\ \nonumber &=&
U<n_{i,-\sigma}d_{i,\sigma}c^{\dagger}_{p,\sigma}>\\
\nonumber &+&\sum_{p'} v_{i,p'}G^{r}_{p',p}+
\sum_{p^{''},p^{'},\sigma} g_{p^{''},p^{'}}
<c^{\dagger}_{p^{''},\sigma'}c_{p^{'},\sigma'}d_{i,\sigma}c^{\dagger}_{p,\sigma}>,
\end{eqnarray}

\begin{eqnarray}
& &(\epsilon-\epsilon_{p',\lambda}-U(N_c-N_\lambda)) G^{r}_{p',p}(\epsilon)\\
\nonumber &=& \delta_{p',p}+v_{p',i}G^{r}_{i,p}+
\sum_{i,p^{''},\sigma}g_{p^{''},p^{'}}
<(n_{i,\uparrow}+n_{i,\downarrow})d_{p^{''},\sigma}c^{\dagger}_{p,\sigma}>,
\end{eqnarray}
and
\begin{eqnarray}
& &(\epsilon-\epsilon_{p',\lambda}-U(N_c-N_\lambda)) G^{r}_{p',j}(\epsilon)\\
\nonumber &=&
v_{p',i}G^{r}_{i,p}+\sum_{i,p^{''},\sigma}g_{p^{''},p^{'}}
<(n_{i,\uparrow}+n_{i,\downarrow})d_{p^{''},\sigma}d^{\dagger}_{j,\sigma}>.
\end{eqnarray}

Here, $\Gamma=(\Gamma_{\ell,L}+\Gamma_{\ell,R})/2$ and
$g_{p,p^{'}}=\frac{U_{dc}}{N}e^{i({\bf p-p}')\cdot {\bf R}_i}$. In
Eqs. (4)-(7), we have introduced four one-particle Green's
functions
$G^{r}_{i,j}(\epsilon)=<d_{i,\sigma}d^{\dagger}_{j,\sigma}>$,
$G^{r}_{i,p}(\epsilon)=<d_{i,\sigma}c^{\dagger}_{p,\sigma}> $,
$G^{r}_{p',p}(\epsilon)=<c_{p',\sigma}c^{\dagger}_{p,\sigma}>$ and
$G^{r}_{p',j}(\epsilon)=<c_{p',\sigma}d^{\dagger}_{j,\sigma}>$.
These four single-particle Green's function are coupled with
two-particle Green's functions via $U$ and $U_{dc}$. The equation
of motion for the two-particle Green's function (defined as
$<n_{i,-\sigma}d_{i,\sigma}d^{\dagger}_{j,\sigma}> $,
$<n_{i,-\sigma}d_{i,\sigma}c^{\dagger}_{p,\sigma}> $,
$<c^{\dagger}_{p^{''},\sigma'}c_{p^{'},\sigma'}d_{i,\sigma}d^{\dagger}_{j,\sigma}>$,
$<c^{\dagger}_{p^{''},\sigma'}c_{p^{'},\sigma'}d_{i,\sigma}c^{\dagger}_{p,\sigma}>$,
$<(n_{i,\uparrow}+n_{i,\downarrow})d_{p^{''},\sigma}c^{\dagger}_{p,\sigma}>$,
and
$<(n_{i,\uparrow}+n_{i,\downarrow})d_{p^{''},\sigma}d^{\dagger}_{j,\sigma}>$)
are coupled to the three-particle Green's functions. In order to
terminate the heirachy of the equation of motions, we use the
Hartree-Fock approximation method$^{19-21}$ to decouple terms
involving the $U_{dc}$ factor. Meanwhile in the derivation for
$<n_{i,-\sigma}d_{i,\sigma}d^{\dagger}_{j,\sigma}> $ and
$<n_{i,-\sigma}d_{i,\sigma}c^{\dagger}_{p,\sigma}> $, the
treatment for coupling terms between localized states and the
electrodes (or 2-D conduction band) is employed in the scheme
considered in our previous method, which is valid for the Coulomb
blockade regime.$^{16,17}$  Solving Eqs. (4)-(7), we obtain

\begin{eqnarray}
G^{r}_{pp'\lambda}(\epsilon)=\frac{\delta_{p,p'}}{\epsilon-\epsilon_p-\Delta_\lambda}
+\frac{v^2
G^{r}_{\ell,\ell}(\omega)}{(\epsilon-\epsilon_p-\Delta_\lambda)
(\epsilon-\epsilon_{p^{'}}-\Delta_\lambda)},
\end{eqnarray}
where $\Delta_\lambda=U_{dc}(N_{d,\sigma}+N_{d,-\sigma})+U(N_c-N_\lambda)$ and
\begin{eqnarray}
G^{r}_{\ell,\ell}(\epsilon)&=&\frac{1-N_{d,-\sigma}}{\epsilon-E_0-\Delta_c-(\Gamma_b(\epsilon)-i\Gamma)}\\
\nonumber
&+&\frac{N_{d,-\sigma}}{\epsilon-E_0-U-\Delta_c-(\Gamma_b(\epsilon)-i\Gamma)}.
\end{eqnarray}
The retarded Green's function $G^{r}_{\ell,\ell}(\epsilon)$ has
the self-energies $\Delta_c=U_{dc} N_c$ and
$\Gamma_b(\epsilon)=\frac{1}{N}\sum_p
\frac{v^2}{\epsilon-\epsilon_p-\Delta_d+i\delta}$ ($\delta$ is a positive infinitesimal number), which results from the
interaction between the localized states and conduction band.
$N_d$  is the occupation number of  $p_z $ orbital in each unit cell.
The second term in Eq. (8) describes the scattering amplitude of the
conduction electron  due to interaction with the $p_z$ orbitals.
This term would be important for studying the charge transport through the $p_x$ and $p_y$ orbitals in the x-y plane .
In this study we focus on the longitudinal transport (along z-axis) rather than transverse transport (in the x-y plane),
thus this term can be ignored.

To reveal the tunneling current behavior, the occupation number
$N_{d,\sigma}$ determining the probability amplitude of resonant
channels $\epsilon=E_0+\Delta_c+(\Gamma_b-i\Gamma) $ and
$\epsilon=E_0+U+\Delta_c+(\Gamma_b-i\Gamma) $  is solved by the equation
\begin{equation}
N_{d,\sigma}=-\int \frac{d\epsilon}{\pi} \frac{\Gamma_L
f_L(\epsilon)+\Gamma_R f_R(\epsilon)}{\Gamma_L+\Gamma_R}
ImG^{r}_{\ell,\ell}(\epsilon).
\end{equation}
As for $N_c$, we have
\begin{equation}
N_{c}=-\sum_{p,\lambda} \int \frac{d\epsilon}{\pi}
\frac{\Gamma_{L,c} f_L(\epsilon)+\Gamma_{R,c}
f_R(\epsilon)}{\Gamma_{L,c}+\Gamma_{R,c}} Im{\cal
G}^{r}_{p\lambda,p\lambda}(\epsilon)/N,
\end{equation}
where ${\cal
G}^{r}_{p\lambda,p\lambda}(\epsilon)=1/(\epsilon-\epsilon_{p,\lambda}-U
(N_c-N_\lambda)-2U_{dc}N_d+i(\Gamma_{L,c}+\Gamma_{R,c})/2)$.  As
mentioned above,  the coupling between the electrodes and the
$p_x/p_y$ orbitals of QD array is negligible. (i.e
$\Gamma_c=(\Gamma_{L,c}+\Gamma_{R,c})/2$, where
$\Gamma_{L,c}(\Gamma_{R,c})$ denotes the tunneling between the
left (right) electrode and $p_x/p_y$ orbitals is small),
therefore, $Im{\cal G}^{r}_{p\lambda,p\lambda}(\epsilon)\approx
\pi \delta(\epsilon-\epsilon_{p,\lambda}-U
(N_c-N_\lambda)-2U_{dc}N_d)$.

The range of applied bias considered here would not be enough to overcome the charging
energy of $U+\Delta_c$, therefore, the second term in Eq. (9) can
be ignored and we have
$G^{r}_{\ell,\ell}(\epsilon)=(1-N_{d,-\sigma})/(\epsilon-E_0-\Delta_c
-\Gamma_b+i\Gamma)$ in which $\Gamma_b(\epsilon)=-i\Gamma_0$ (ignoring the small real part).  The
occupation number at zero temperature is calculated by
\begin{eqnarray}
N_{d,\sigma}&=&\frac{(1-N_{d,-\sigma})}{\pi}
\frac{\Gamma_L}{\Gamma_L+\Gamma_R} \\ \nonumber&
&\int_{-\infty}^{E_F+eV_a} d\epsilon \frac{\Gamma_0+\Gamma}
{(\epsilon-E_0-\alpha eV_a-\Delta_c)^2+(\Gamma_0+\Gamma)^2}
\end{eqnarray}
or
\begin{eqnarray}
& &\frac{\Gamma_L+\Gamma_R}{\Gamma_L}\pi N_{d}/(1-N_d)\\
\nonumber &=&\cot^{-1}(\frac{E_F+eV_a-E_0-\alpha eV_a-U_{cd}
N_c}{\Gamma_0+\Gamma}),
\end{eqnarray}
in which $\alpha e V_a$ term arises from the applied bias crossing
QDA and $\alpha$ is a dimensionless scaling factor determined by
the QDA location, and \be N_{\lambda}=
\frac{\Gamma_{L,c}}{\Gamma_{L,c}+\Gamma_{R,c}}\int_{-\infty}^{E_F+eV_a}
d\epsilon D_\lambda(\epsilon-2U_{dc}N_d-U(N_c-N_\lambda)), \ee
where
$D_\lambda(\epsilon)=\sum_{p}\delta(\epsilon-\epsilon_{p,\lambda})/N$
denotes the density of states per unit cell of the $\lambda$-th
conduction band. Due to the fact that the $p_z$ energy level is
always above the Fermi energy of right electrodes (in the range of
bias considered), we can ignore the electron injection from the
right electrode in Eqs. (12) and (14). We first consider the
simple case in which  $N_x=N_y$ (valid for a square lattice) and
we approximate the density of states by a square pulse function
\be D_x(\epsilon)= D_y(\epsilon)=1/W \mbox{ for } E_b < \epsilon
<E_b + W, \ee where $E_b$ denotes the bottom of the conduction
band and $W$ is the band width. Such an approximation allows Eq.
(11) to have a simple analytic solution of the form
\begin{equation}
N_{\lambda} = b - c N_d
\end{equation} with $b=[E_f+(1-\alpha)eV_a-E_b]/(\gamma W+3U)$ and
$c=2U_{dc}/(\gamma W+3U)$, where
$\gamma=(\Gamma_{L,c}+\Gamma_{R,c})/\Gamma_{L,c}$. Substituting
this into Eq. (13) allows a simple transcendental equation, which
can be solved numerically. The equation allows a maximum of three roots, out of which only two are stable roots.

We can also solve two coupled transcendental equations as given in
Eqs. (13) and (14) numerically for a more realistic density
states, which is derived for a 2D tight-binding model. We consider
a tight-binding model for $p_x$ and $p_y$ orbitals arranged on a
rectangular lattice with lattice constants $a$ and $b$. Figure 2
illustrates the rectangular lattice.  The band structure for the
$p_x$ band is given by \be E_x({\bf k}) = E_p - 2 v_l \cos (k_x a)
- 2v_t \cos (k_y b), \ee where $v_l$ denotes the $(pp\sigma)$
interaction and $v_t$ denotes the $(pp\pi)$ interaction.$^{24}$
For the $p_y$ band, we have \be E_y({\bf k}) = E_p - 2 v'_l \cos
(k_y b) - 2v'_t \cos (k_x a). \ee The density of states per unit
cell form the $p_x$ band  is given by (if $v_l>v_t$)
\be D_x(\epsilon) = \left\{ \begin{array}{lll} \frac 1{\pi^2} \int_0^{\pi}  d\eta [(2v_l)^2-(2v_l+2v_t(1-\cos \eta) -\tilde \epsilon)^2]^{-1/2}\theta(\tilde \epsilon-2v_t(1-\cos \eta) ) \; \mbox{ for } 0< \tilde\epsilon <  4 v_t \\
\frac 1{\pi^2} \int_0^{\pi}  d\eta  [(2v_t)^2-(\tilde \epsilon-2v_t-2v_l(1-\cos \eta))^2 ]^{-1/2}\theta(\tilde \epsilon-2v_l(1-\cos \eta) ) \; \mbox{ for }  4 v_t < \tilde \epsilon <  4 v_l \\
\frac 1{\pi^2} \int_0^{\pi}  d\eta [(2v_l)^2-(2v_l+2v_t(1+\cos
\eta) -\bar \epsilon)^2]^{-1/2}\theta(\bar\epsilon- 2v_t(1+\cos
\eta) ) \; \mbox{ for } 0< \bar\epsilon <  4 v_t \end{array}
\right. , \ee where $\tilde\epsilon = \epsilon -E_p+2v_l+2v_t$ and
$\bar \epsilon = E_p+2v_l+2v_t -\epsilon$. If $v_l<v_t$, then the roles of $v_l$ and $v_t$
should be exchanged in the above expression. The DOS described by Eq. (19) contains
the Van Hove singularities. Similar expression
($D_y(\epsilon)$) holds for the $p_y$ band with the hopping
parameters $v_l$ and $v_t=v^{'}_t$ replaced by
$v'_l=v_l$ and $v'_t$. By varying these hopping parameters
(for instance, fix lattice constant a and tune b), we can study
the behavior of bistable tunneling current for systems between the
1D and 2D limits.

We numerically solve the coupled nonlinear Eqs. (13) and (14) for
$U_{dc}=U=50 meV$, $\Gamma_L=1~meV (\Gamma_{L,c}=\Gamma_L/10)$,
and $\Gamma_R=1 meV(\Gamma_{R,c}=\Gamma_R/10)$.  Throughout the
paper, we shall use $T=0 K$, $v'_t=5 meV$, $v_l=20 meV$,
$\alpha=0.5$, and $E_F+V_0 = E_p$, where $V_0$ is a reference bias
for $V_a$. For simplicity, $v_0=0$. The tight-binding parameters
are assumed to scale according to the $1/R^2$ rule$^{24}$, where
$R$ is the separation between two QDs. Thus, we have $v'_l=v_l
(a/b)^2 $ and $v_t=v^{'}_t (a/b)^2$. The occupation number $N_d$
as a function of the applied bias at zero temperature for the
square lattice case ($a=b$) is shown as solid line in Fig. 3. The
result remains very similar if we use the constant DOS
approximation as described in Eqs. (15) and (16) with the same
bandwidth, $W=4(v_l+v_t)$. Although there are Van Hove
singularities in tight binding DOS, the structure of
bias-dependent occupation number does not exhibit an anormal
feature. This is because $p_z$ orbital is correlated with $p_x$
and $p_y$ via $N_c$, which is related to the integral over the
DOS.

We see that the occupation number, $N_d$ has bistable roots. Although, the
QDA crystal structure reported in references[10,13] has a
triangle lattice, the results of Fig. 3 indicate
that the hysteresis behavior will not depend on the
detailed band structure. Once the occupation numbers are solved, the
tunneling current can be obtained by $J=\frac{e}{\hbar}\Gamma_R
N_d=J_0 N_d$, which is valid for zero temperature and when the carrier injection from the right lead can be ignored. Consequently, $N_d$ via
the applied bias directly shows the tunneling current
characteristics. The roots for
the turn-on and turn-off processes in Fig. 3 are determined by
selecting the root closest to the root corresponding to the previous value of
$V_a$ when multiple roots are allowed.

It is crucial to clarify how the system selects the high
conductivity state (larger $N_d$) or the low conductivity state
(smaller $N_d$) as the applied bias is turned on or off. In Fig.
4a, we plot the bistable current for various strengths of
$U_{dc}$. Curves 1, 2, and 3 denote, respectively, $U$=50, 30, and
20 $~meV$. For smaller Coulomb interactions, the bistable current
vanishes. The critical Coulomb interactions to maintain the
bistable current depend on the physical parameters such a
bandwidth of 2-D conduction band, tunneling rates between dot
array and electrodes, the broadening $\Gamma_0$, and temperatures.
From the application of memory devices, larger $U$ (smaller dot
size) is favored because the bistability behavior will be more
robust against the increase of temperature and broadening.
Although Fig. (4a) exhibits the bistable current, it does not show
any negative differential conductivity (NDC), unlike the bistable
current in quantum well systems, which is typically associated
with NDC.$^{1,2}$ Fig. 4b shows the behavior of $N_c$ for the same
set of on-site Coulomb interaction strengths as in Fig. 4a. It is
useful to understand the  behavior of $N_c$ for clarifying the
bistable mechanism of $N_d$, which is described below.

The physical mechanism for this bistablity behavior can be
explained as follows. As we gradually increase the bias from below
the resonance level $E_p$, the allowed solution to the occupation
number, $N_d$ remains small during the "turn-on" process, as a
result of inter-level Coulomb blockade (i.e.  the $p_z$ level is
pushed up by the amount $N_cU_{dc}$,  where the 2-D conduction
band occupation number $N_c$ is appreciable). Once the applied
bias reaches a critical value (the inter-level Coulomb blockade is
overcome), $N_d$ increases quickly to a value around 1/3. Now,
charges accumulate in the localized $p_z$ orbitals, which leads to
an increase of the self-energy of the conduction band states by
$2U_{dc}N_d$, and hence reduces $N_c$ to a much smaller value
(from $N_{c2}$ to $N_{c3}$ as illustrated in Fig. 4b). This in
turn causes the $p_z$ energy level to decease during the
"turn-off" process while $N_d$ maintains around 1/3 to keep the
transcendental equation self-consistent. When the applied bias
continues to decrease to a critical value, $N_c$ is switched from
$N_{c4}$ to $N_{c1}$ (see Fig. 4b), and $N_d$ quickly goes back to
the lower value which becomes the only allowed self-consistent
solution to the transcendental equation. The results of Fig. 4
demonstrate that the bistable tunneling current arises from the
on-site repulsive Coulomb interactions $U_{dc}$. In the polaron
model adopted in Refs. 6 and 7, the bistable mechanism arises from
an attractive potential of $-\frac{2\lambda_p^2}{W_0}
d^{\dagger}_{\ell,\sigma} d_{\ell,\sigma}
d^{\dagger}_{\ell,-\sigma} d_{\ell,-\sigma}$,where $\lambda_p$ and
$W_0$ are the electron-phonon interaction strength and the phonon
frequency. This two particle interaction term should be corrected
as $(U-\frac{2\lambda_p^2}{W_0}) d^{\dagger}_{\ell,\sigma}
d_{\ell,\sigma} d^{\dagger}_{\ell,-\sigma} d_{\ell,-\sigma}$ when
the intralevel Coulomb interaction is included in a single QD or
molecule. Due to the large repulsive intralevel Coulomb
interaction U, a net attractive electron-electron interactions
mediated by phonon is difficult to achieve, which may explain why
a bistable tunneling current through a single QD junction has not
been observed.

To realize nanoscale memory structures, we need to examine whether
this bistable current exists in one dimensional array. Figure. 5
show tunneling current through $p_z$ orbital for different ratios
of $b/a$. Other parameters are the same as those for Fig. 3. From
two dimensional array to quasi-one dimensional array, the
bistability behavior is sustained (although somewhat weaker in the
1D limit). When $b/a$ is greater than 3, (results not shown here)
the bistable current behaves  essentially the same as in the
$b/a=3$ case. This indicates that we already reached the 1D limit
for $b/a \approx 3$, and the bistable current still exists. If we
assume $a=3 nm$,  $b=9 nm$, and 50 coupled QDs along the chain in
the $x$ direction are needed to establish a band-like behavior,
then the density of the memory device is around $1/(1350
nm^2)\approx 0.5TB/in^2$. The operating voltage needed is around
100mV, which indicates very low power consumption. Because this is
a quantum device, the switching time is expected to be comparable
to the tunneling rate, which is on the order of 1 THz.

Finally, we show in Fig. 6 the tunneling current for various
tunneling rate ratios $\Gamma_L/\Gamma_R$ for the $b/a=2$ case. It
is seen that the bistability disappears for $\Gamma_L=0.1$~meV and
$\Gamma_R=1~$meV (shell-tunneling condition). In this case,
charges are unable to accumulate in the $p_z$ orbitals.
Consequently, the effect of $U_{dc}$ is suppressed. On the other
hand, for $\Gamma_L=1$~meV and $\Gamma_R=0.1$meV (shell-filling
condition), the effect of $U_{dc}$ is enhanced. This leads to the
wider voltage range of bistable current. From results of Fig. 6,
we can control the bistable current by adjusting the tunneling
rate ratio.


In conclusion, we have illustrated a novel mechanism for
generating the bistable tunneling by using a junction involving a
2D periodic array of QDs with six-fold degenerate $p$-like
states, which consist of localized $p_z$ orbitals interacting with
non-localized $p_x$ and $p_y$ orbitals via the on-site Coulomb
interaction. Due to the interplay of Coulomb blockade effect for
the localized state and the self-energy correction to the 2D
conduction bands formed by the $p_x$ and $p_y$ orbitals, a
bistable tunneling current with well defined hysteresis behavior
can be achieved. This bistable current is not sensitive to
the details of the band structure, but sensitive to the charging energy and
the ratio of incoming to outgoing tunneling rate. Such a hysteresis behavior arises from
a collective effect, not observable in a single QD. It is shown
that this hysteresis behavior can also exist in one dimensional QDA  and
very high density integrated memory circuits can be realized.


This work was supported by the National Science Council of the
Republic of China under Contract Nos. NSC 97-2112-M-008-017-MY2,
96-2120-M-008-001, and 95-2112-M-001-068-MY3.




\mbox{}

\newpage

{\bf Figure Captions}

Fig. 1. Quantum dot array is embedded into an insulated matrix
sandwiched between two metallic electrodes.

Fig. 2. Two band ($p_x$ and $p_y$) rectangular lattice with
lattice constants $a$ and $b$. $v_l (v^{'}_t)$ and $v_t (v^{'}_l)$
denote, respectively, the electron hopping strength of $p_x (p_y)$
orbital in $a$ and $b$.

Fig. 3. Occupation number as a function of applied bias for
$U_{dc}=U=50 meV$, $\Gamma_L=1~meV(\Gamma_{L,c}=\Gamma_L/10)$, and
$\Gamma_R=1~meV(\Gamma_{R,c}=\Gamma_R/10)$. Solid line and dashed
line denote, respectively, tight binding DOS and constant DOS.

Fig. (4a) Bistable current as a function of applied bias for
variation strengths  $U$ at zero temperature.  Fig. (4b) shows
$N_c$ for different $U$. Other parameters are the same those of
Fig. 3. Note that tunneling current is in units of
$J_0=e\Gamma_R/\hbar$, which depends on the tunneling rate of
$\Gamma_R$.

Fig. 5. Bistable current as a function of applied bias for $b/a=$
1, 2, and 3. Other parameters are the same those for Fig. 3.

Fig. 6. Bistable current as a function of applied bias for the
various ratios of $\Gamma_L/\Gamma_R$ and $b/a=3$. Other
parameters are the same as those for Fig. 3.
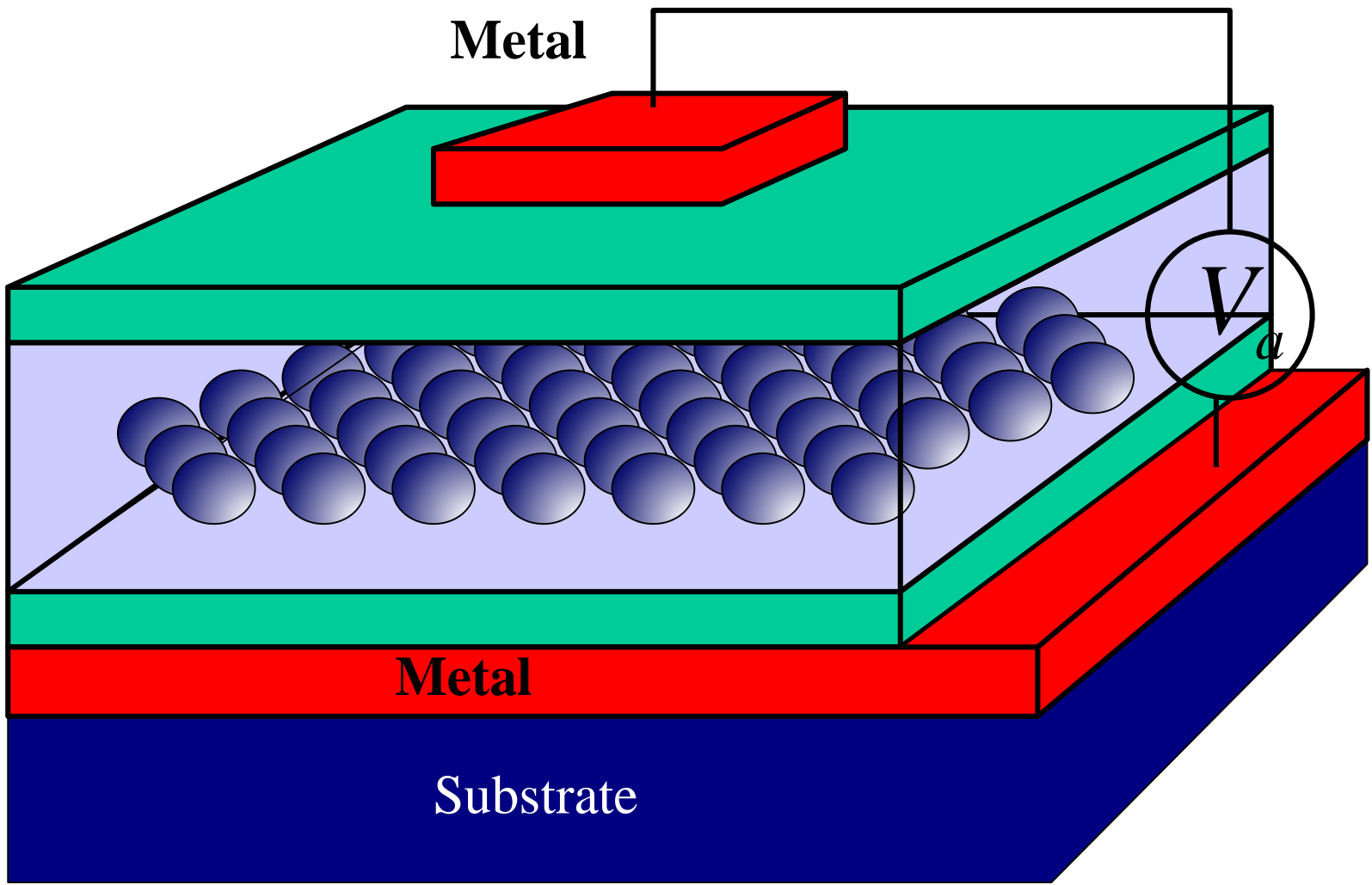
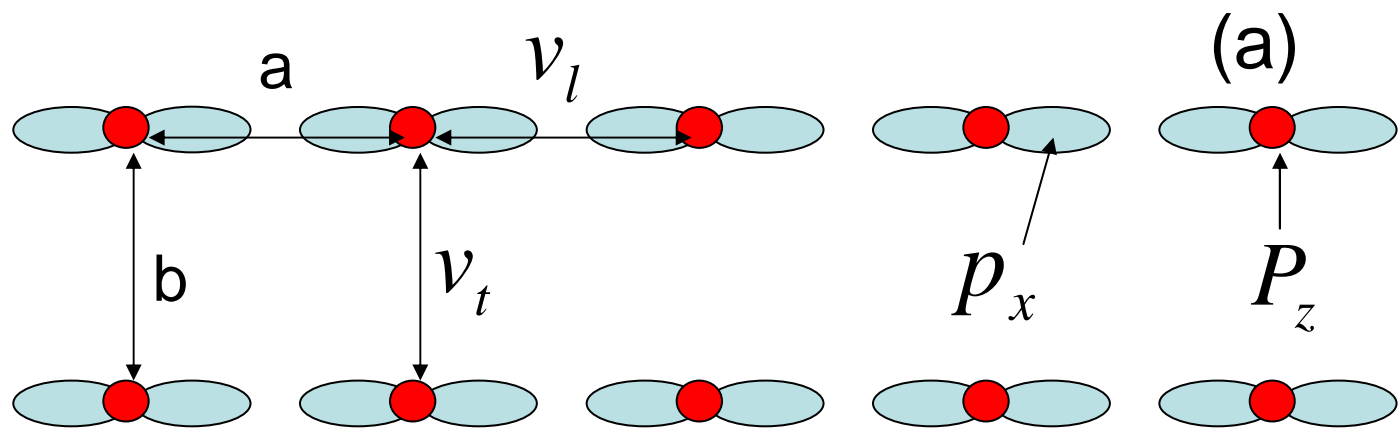
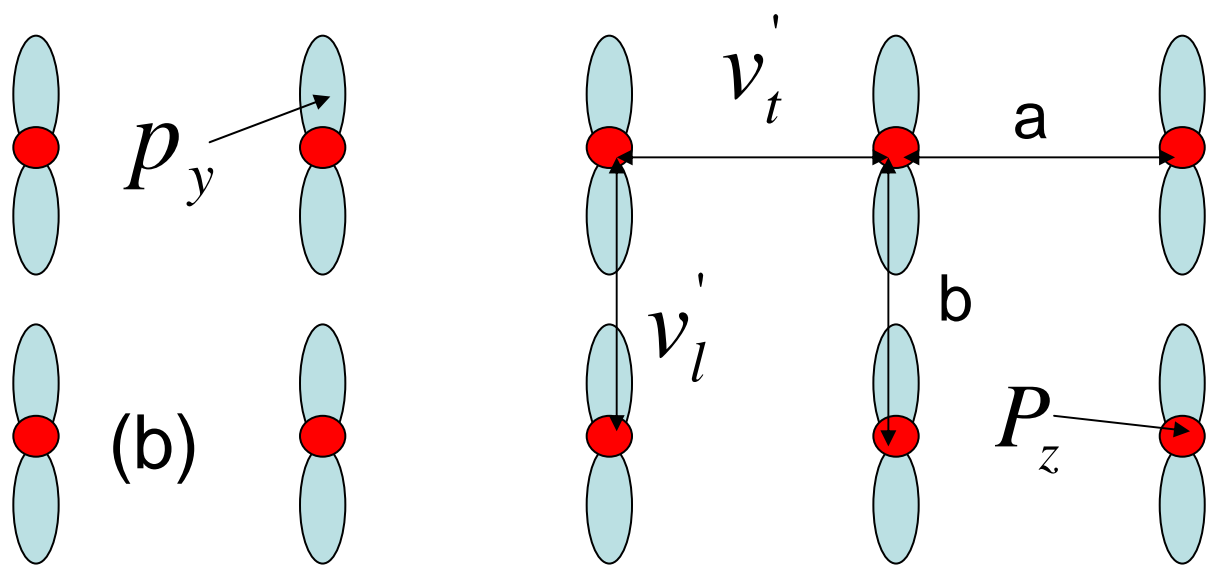
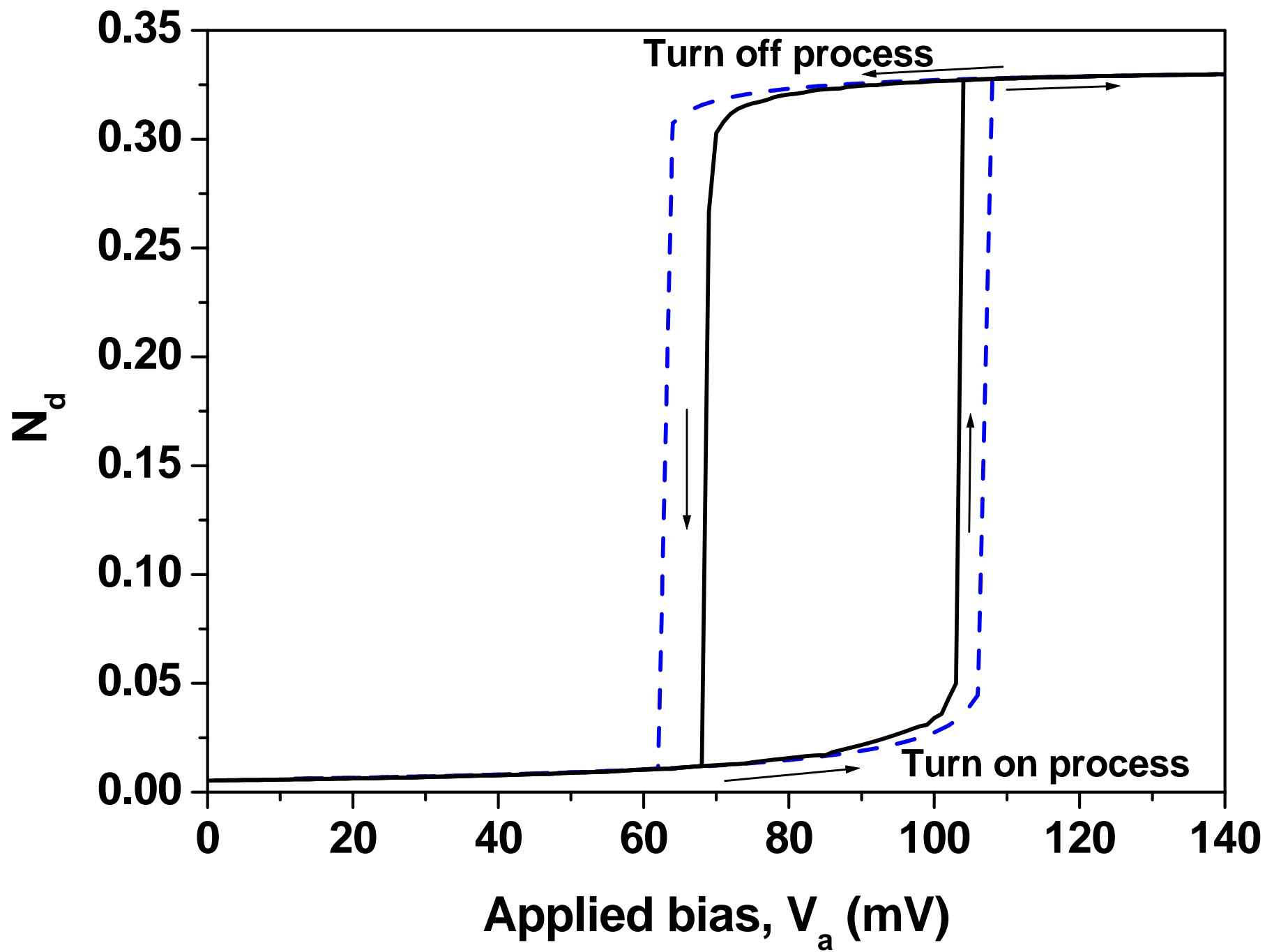
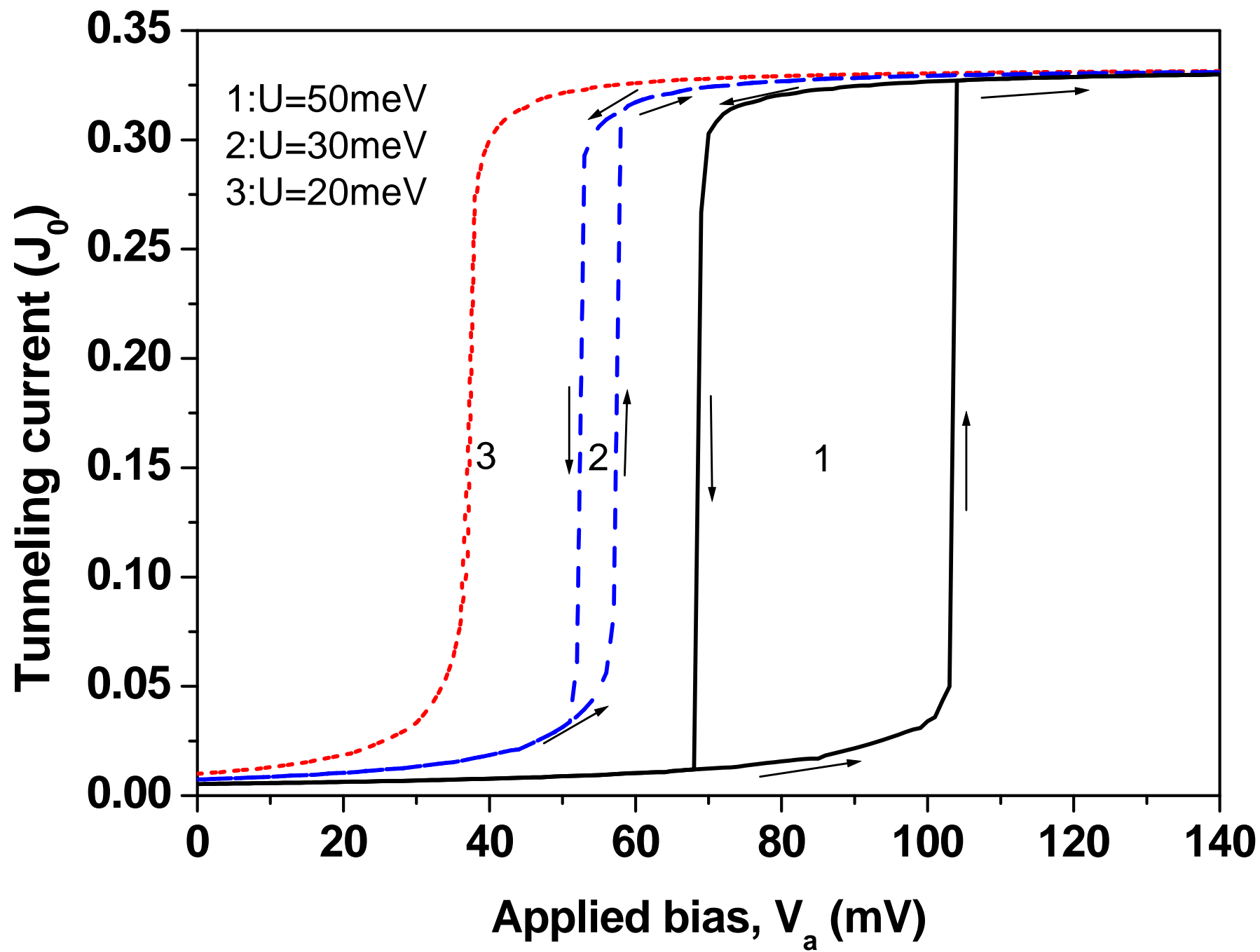
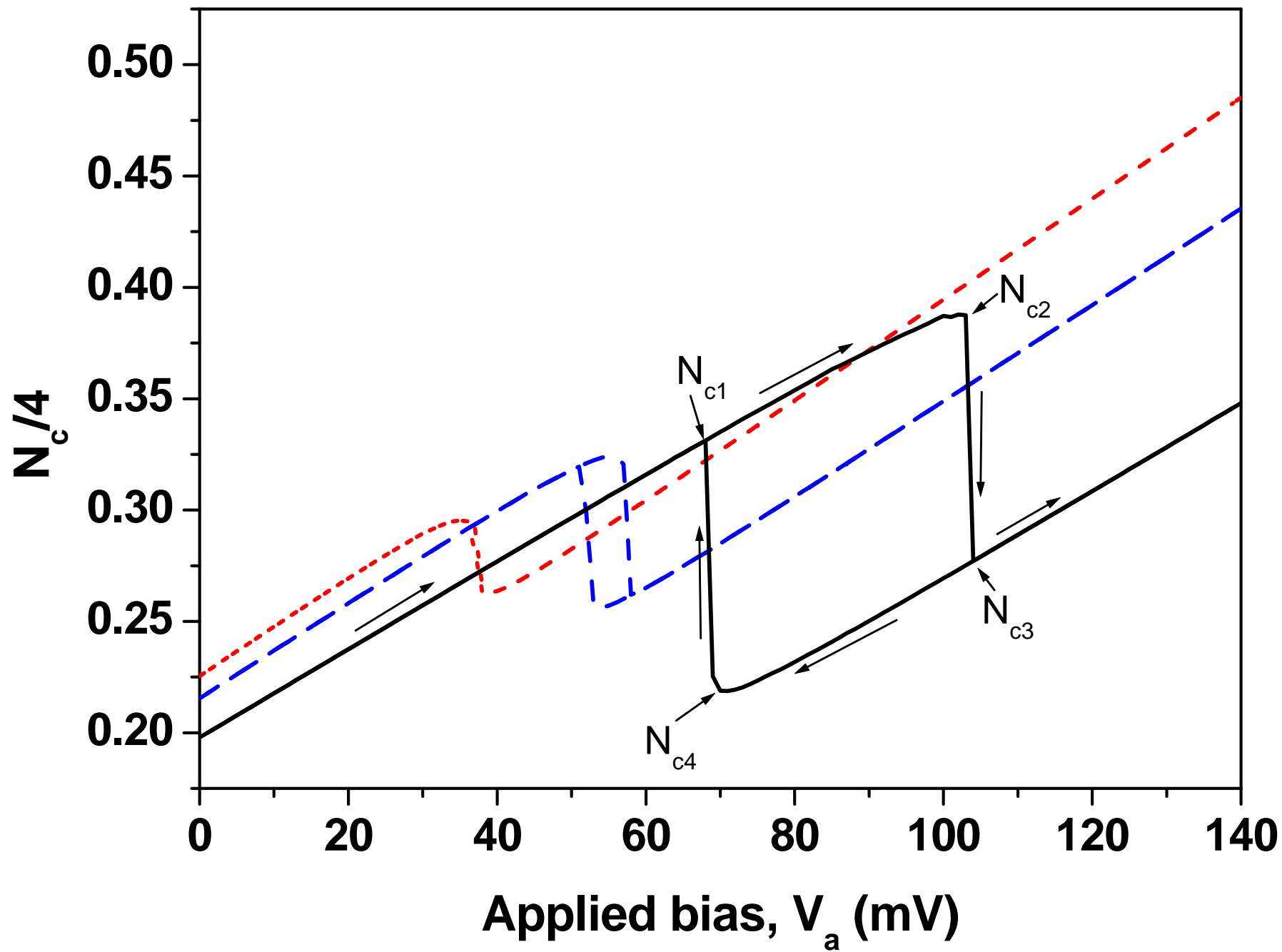
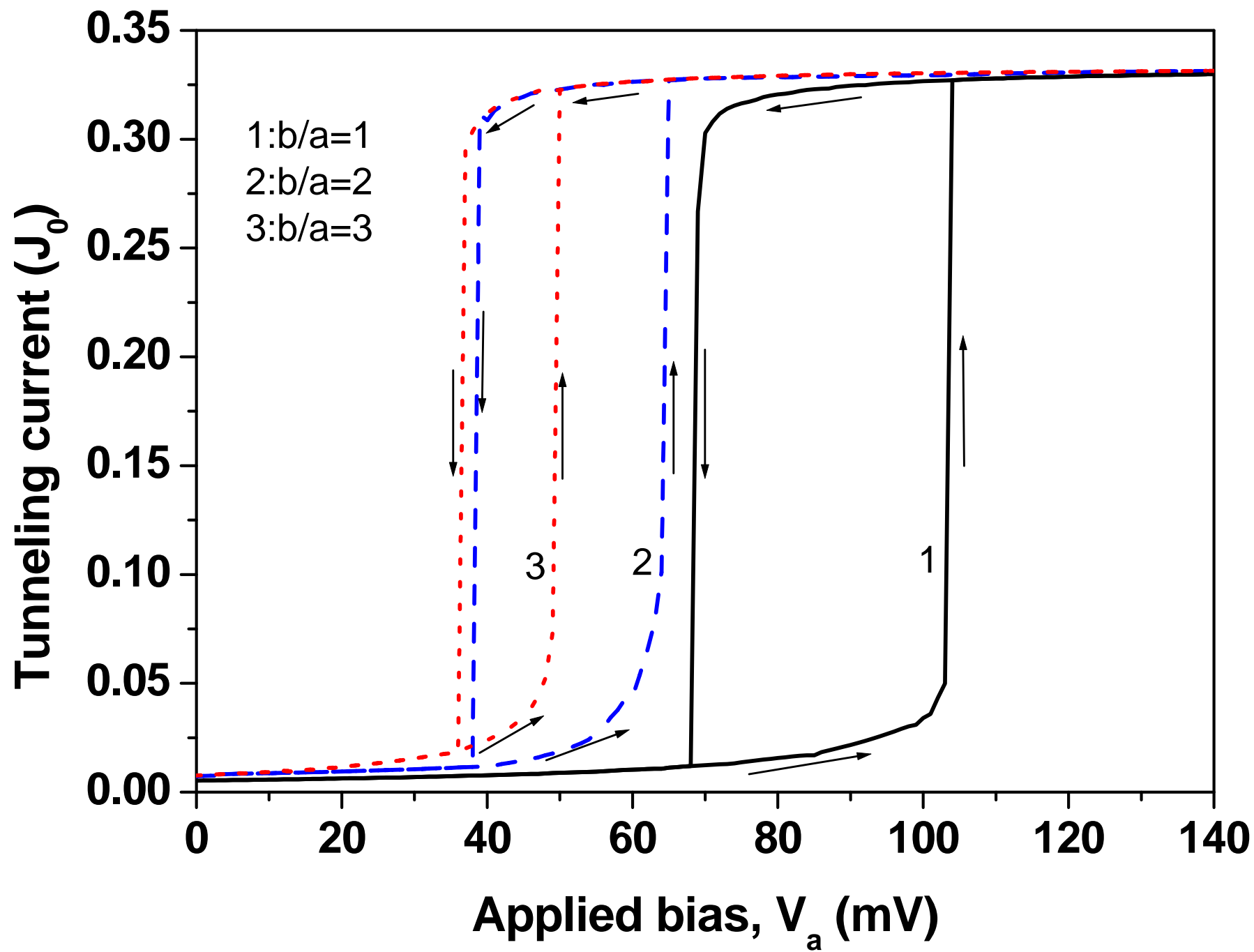
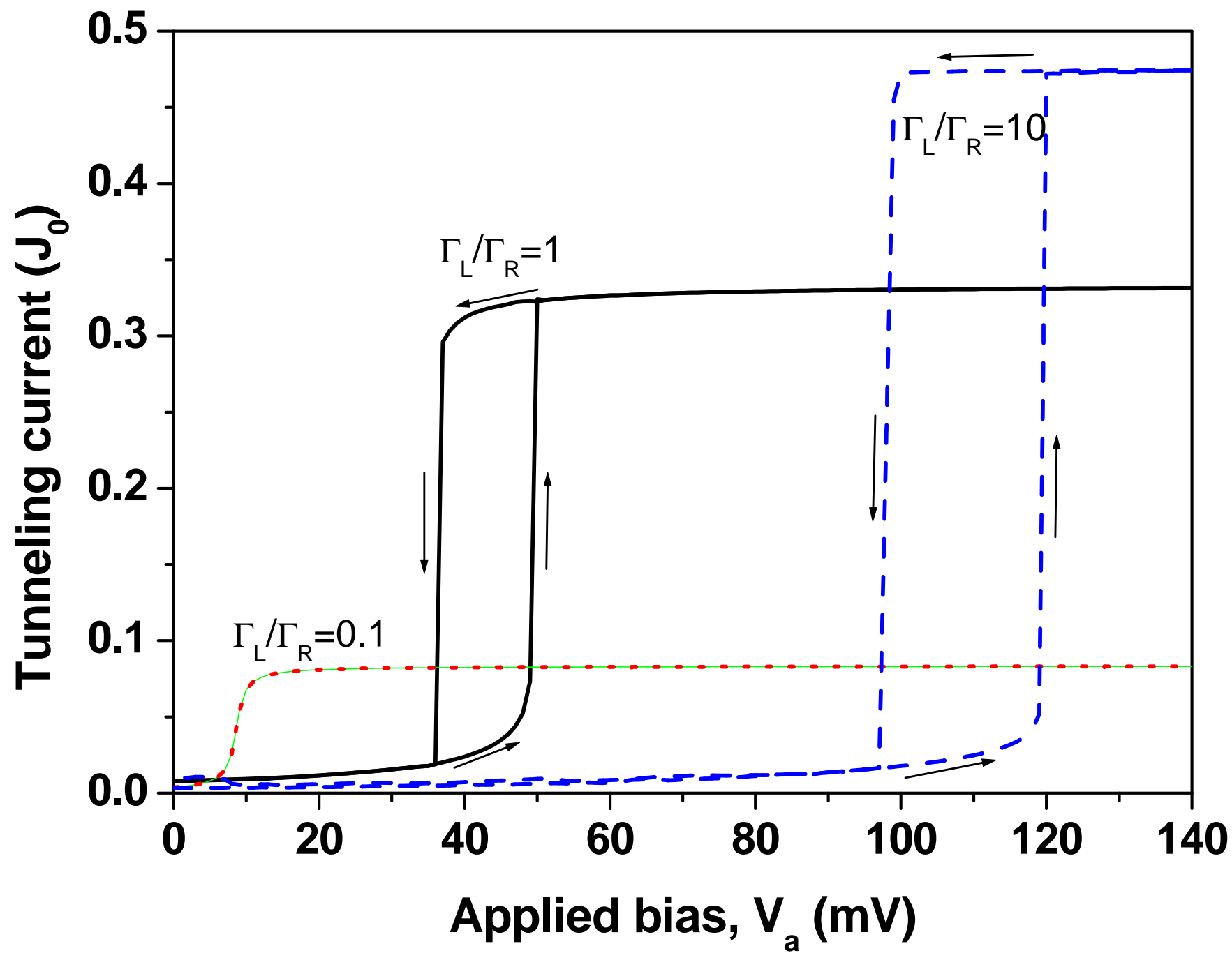
\end{document}